\documentclass[11pt,a4paper]{article}
\usepackage[T1]{fontenc}
\usepackage[utf8]{inputenc}
\usepackage{authblk}
\usepackage{graphicx}
\usepackage{mathtools}
  \usepackage{dsfont}
  \usepackage{amsmath}
  \usepackage{amsthm}
  \usepackage{amssymb}

  \theoremstyle{definition}
 \newtheorem{defn}{Definition}
  
  \newtheorem*{defn*}{Definition}
 
  \theoremstyle{plain}
  \newtheorem{thm}{Theorem}
  
  \newtheorem{prop}{Proposition}
  \newtheorem*{prop*}{Proposition}
  \newtheorem{lem}{Lemma}
   \newtheorem*{cor*}{Corollary}
  \newtheorem*{theo*}{Theorem}
  \newtheorem*{thm*}{Theorem}
  
  \theoremstyle{remark}
  \newtheorem{rem}{Remark}
 
 \usepackage[toc,page]{appendix}
\usepackage{calrsfs}
\usepackage{tikz-cd}

\newcommand{\p}{\mathbb{P}}

\newcommand{\R}{\mathbb{R}}
\newcommand{\A}{\mathcal{A}}

\newcommand{\C}{\mathcal{C}}

\newcommand{\G}{\mathcal{G}}

\usepackage{pstricks,pst-node,pst-tree}

\usepackage{bm}

\usepackage{relsize}

 \usepackage{xcolor}

 \newcommand\cat[1]{\textbf{#1}}

\newcommand{\Pa}{\mathcal{P}}

\newcommand{\mes}{\cat{Mes}}
\newcommand{\Kern}{\cat{Kern}}
\newcommand{\Vect}{\cat{Vect}}
\newcommand{\Mod}{\cat{Mod}}
\newcommand{\dd}{\text{d}}

\newcommand{\Set}{\cat{Set}}

\newcommand{\colim}{\operatorname{colim}}

\makeatletter
\newtheorem*{rep@theorem}{\rep@title}
\newcommand{\newreptheorem}[2]{%
\newenvironment{rep#1}[1]{%
 \def\rep@title{#2 \ref{##1}}%
 \begin{rep@theorem}}%
 {\end{rep@theorem}}}
\makeatother

 \theoremstyle{plain}
\newreptheorem{thm}{Theorem}
\newreptheorem{prop}{Proposition}
\newreptheorem{cor}{Corollary}

\usepackage{comment}

\makeatletter
 \let\Ginclude@graphics\@org@Ginclude@graphics
\makeatother
\newcommand\imCMsym[4][\mathord]{%
  \DeclareFontFamily{U} {#2}{}
  \DeclareFontShape{U}{#2}{m}{n}{
    <-6> #25
    <6-7> #26
    <7-8> #27
    <8-9> #28
    <9-10> #29
    <10-12> #210
    <12-> #212}{}
  \DeclareSymbolFont{CM#2} {U} {#2}{m}{n}
  \DeclareMathSymbol{#4}{#1}{CM#2}{#3}
}
\newcommand\alsoimCMsym[4][\mathord]{\DeclareMathSymbol{#4}{#1}{CM#2}{#3}}

\imCMsym{cmmi}{124}{\CMjmath}
\imCMsym[\mathop]{cmsy}{113}{\CMamalg}
\imCMsym[\mathop]{cmex}{96}{\CMcoprod}
\alsoimCMsym[\mathop]{cmex}{97}{\CMbigcoprod}

\newcommand{\id}{\operatorname{id}}

\makeatletter

  \theoremstyle{plain}

\makeatletter
\newcommand{\mylabel}[2]{#2\def\@currentlabel{#2}\label{#1}}
\makeatother

\numberwithin{equation}{section}

\usepackage{multiaudience}
\SetNewAudience{long}
\SetNewAudience{GSP}

\newcommand{\Pro}{\mathbb{P}}

\usepackage{minitoc}  
 
\usepackage{changepage} 
 
\usepackage{chngcntr}
\usepackage{etoolbox}
 \makeatletter
   {\par}
\makeatother

\AtBeginEnvironment{subappendices}{%
\section*{Appendix}
\addcontentsline{toc}{section}{Appendices}
\counterwithin{figure}{section}
\counterwithin{table}{section}
}

\usepackage{cite}
\usepackage{hyperref} 
\usepackage[disable]{todonotes}
\DeclareMathOperator{\Fac}{\operatorname{Fac}}

\begin{document}

\title{Compositional statistical mechanics, entropy and variational inference}
%
%

\author[1]{Grégoire Sergeant-Perthuis\thanks{gregoireserper@gmail.com}}
\affil[1]{Laboratoire de Biologie Computationnelle et Quantitative, Sorbonne Université, Paris , France \\  ORCiD: 0000-0002-2079-3410}

%

%
\maketitle              

 \begin{abstract}
In this document, accepted for the Twelfth Symposium on Compositional Structures (\href{https://www.cl.cam.ac.uk/events/syco/12/}{SYCO 12}) \cite{sergeant-perthuis2024compositional}, we aim to gather various results related to a compositional/categorical approach to \textit{rigorous} Statistical Mechanics \cite{sergeantperthuis2021bayesiangraphoid,sergeantperthuis2021intersectiond,sergeantperthuis2021interactionp,bennequin2020extrafine,sergeantperthuis2021interaction,GSPthese,10.1007/978-3-031-38271-0_26,sergeantperthuis:hal-04456412}. Rigorous Statistical Mechanics is centered on the mathematical study of statistical systems. Central concepts in this field have a natural expression in terms of diagrams in a category that couples measurable maps and Markov kernels \cite{10.1007/978-3-031-38271-0_26}. We showed that statistical systems are particular representations of partially ordered sets (posets), that we call $\mathcal{A}$-specifications, and expressed their phases, i.e., Gibbs measures, as invariants of these representations. It opens the way to the use of homological algebra to compute phases of statistical systems. Two central results of rigorous Statistical Mechanics are, firstly, the characterization of extreme Gibbs measure as it relates to the zero–one law for extreme Gibbs measures, and, secondly, their variational principle which states that for translation invariant Hamiltonians, Gibbs measures are the minima of the Gibbs free energy. We showed in \cite{sergeantperthuis:hal-04456412} how the characterization of extreme Gibbs measures extends to $\mathcal{A}$-specifications; we proposed in \cite{sergeantperthuis2022regionalized} an Entropy functional for $\mathcal{A}$-specifications and gave a message-passing algorithm, that generalized the belief propagation algorithm of graphical models (see O. Peltre's SYCO 8 talk, \cite{PeltrePhD} and \cite{yedidia_constructing_2005}), to minimize variational free energy.
\end{abstract}




\section{Introduction}

\hspace{\parindent} \textbf{Context:} Statistical physics is a framework that focuses on the probabilistic description of complex systems: a collection of interacting `particles' or components of a whole in most generality \cite{doi:10.1142/4090}. Its main feature is to introduce an energy function $H$ for the system, that associates to any of its configurations a real value; the probability $p$ of a configuration is given in terms of the Boltzmann distribution ($p=e^{-\beta H}/\int dx e^{-\beta H(x)} $), which associates high-energy configurations to unlikely events. We will call `statistical system' the complex system provided with the Boltzmann distribution. Statistical physics serves as a rich framework for probabilistic modeling. It has several names depending on the community \cite{928a56b7d6f1473e930f282a0c4b534e}; for example, it is called `energy-based modeling' in machine learning, the probabilistic model is called a Gibbs Random Field in graphical modeling and is a particular case of a statistical system. It is widely used in engineering, for example: computational structural biology (in computational statistical physics: Hamiltonian Monte Carlo) \cite{doi:10.1142/p579,chipot_free_2007}, robotics (reinforcement learning: Markov Chains, Markov Decision Processes) \cite{wiering_reinforcement_2012}, and more generally for modeling interaction and dependencies of random variables using graphical models \cite{Mezard}.

\textbf{Applied Category Theory:} New foundations, based on topology and geometry, were proposed for probability theory, information theory, and deep learning \cite{Gromov,Fritz,Vigneaux2020information,Spivak,Belfiore2021ToposAS}. More generally, they fall in a field of research that has recently emerged, Applied Category Theory, which focuses on applying these principles to engineering \cite{Fritz,Spivak,Fritz1,Marcolli,Baudot2015,Vigneaux2020information}. Recent results give a characterization of the zero–one law for independent random variables and for Markov chains in a categorical formulation \cite{fritz2021finetti,MOSS_PERRONE_2023}. The zero–one law for extreme Gibbs measures is known to extend the ones of independent random variables and Markov chains \cite{Georgii+2011}, so it would be expected that the categorical formulation of extreme Gibbs measures we propose may also relate to the categorical formulation developed in the case of independent random variables and Markov chains.

\textbf{Motivation:} A landmark rigorous formulation of Statistical Mechanics can be found in Georgii's \emph{Gibbs Measures and Phase Transition} \cite{Georgii+2011}. Such a framework, which revolves around the concepts of `specification' and Hamiltonian, is needed to define rigorously pure phases or, more generally, Gibbs measures of statistical systems. For a given Hamiltonian, there can be multiple phases only for infinitely many interacting particles. The main constructions of such a framework rely on the necessity to have a \emph{universe}, denoted $\Omega$, that encompasses all possible configurations of the system, i.e., all the possible joint configurations of the particles $\Omega := \prod_{i \in \mathbb{N}} E_i$, with $E_i$ the state space of the particle $i$. There are important limitations to making references to a `global' universe \cite{Gromov}. Let us state two of them. The first one is that it can be difficult to compute with Hamiltonians (e.g., computing free energies \cite{doi:10.1142/p579}, expectations of observables \cite{alquier_approximate_2020}, infinite volume Gibbs measures \cite{Georgii+2011}) as it requires summing over a set which 'size' increases exponentially with the number of components (variables) involved in the system. The second one is that they don't account for heterogeneity, incompleteness, or incompatibility in the description of the system to model: by definition, each configuration of the system corresponds to observing simultaneously and without loss of information the state of all its components. This second point is related to the first one as to model possible configurations of a system it is required to weight all possible simultaneous configurations of its particles; this forbids granularity of the description of the system: invariably, the dimension of the space of possible models increases exponentially with the number of (elementary) particles. Mathematically, Hamiltonians are sums of local potentials, however, the relation between potentials and their phases is highly nontrivial and noncompositional. In other words, the Hamiltonian point of view does not allow for building complex statistical systems from simple ones in a way that allows controlling and computing the phases of the associated statistical systems from the phases of the simpler one. The relation between the complexity of the interactions that appear in the statistical system and its number of (pure) phases (more than $2$) is known only for very specific systems (e.g. Ising model); results in this direction are considered difficult problems \cite{PhysRev.65.117,PhysRevD.86.025022}. In our approach, we want to forget about the `$\Omega$' and we want to study statistical systems `locally' focusing on their local interactions and how they compose. We propose a constructive approach: `composing' statistical systems together by building in a controlled and computable manner nontrivial statistical systems from simpler ones. This is what reformulating specifications as presheaves allows us to do, in particular, by making use of operations such as coproducts and products but also geometric morphisms. The approach we propose also has the advantage to model statistical systems for which it is not possible to have complete knowledge of the states of the particles at the same time and that could also have conflicting local descriptions inconsistent with a global description.

 \textbf{Contribution:} In this paper, we gather various results related to a compositional/categorical approach to `rigorous' Statistical Mechanics \cite{sergeantperthuis2021bayesiangraphoid,sergeantperthuis2021intersectiond,sergeantperthuis2021interactionp,bennequin2020extrafine,sergeantperthuis2021interaction,GSPthese,10.1007/978-3-031-38271-0_26,sergeantperthuis:hal-04456412}.

Our motivation in \cite{GSPthese} was to build a bridge between geometry and rigorous statistical physics by showing that the standard formal definition of a statistical system, a `specification' (Definition 1.23 \cite{Georgii+2011}), can be identified with a particular representation of a partially ordered set (poset), i.e., a functor of a poset in a certain category. For a given poset $\mathcal{A}$, we call such representation an $\mathcal{A}$-specification. Representations of posets have a precise geometric interpretation \cite{Simson1993LinearRO,https://doi.org/10.48550/arxiv.2010.09651,YANAGAWA2001341,https://doi.org/10.48550/arxiv.2112.02609,incidence}, and there is a rich literature coming from algebra, geometry, and topology to study them. We showed that phases of statistical systems are geometric invariants of these representations and computed them for ‘projective' poset representations. We remarked that, in our setting, systems of ‘finite size' can have multiple phases, which is not possible in the current formalism for phase transition, but we do not claim that those finite size systems capture correctly the possible phase transitions of infinite size systems. However, we believe that the algebraic treatment of statistical systems we propose to be fairly similar for finite size systems and infinite size systems.

In \cite{sergeantperthuis2021interactionp}, we gave a characterization of independent variables in terms of projective objects in the category of presheaves over a poset and an easy-to-verify condition that characterizes such objects. Injective representations were characterized in \cite{sergeantperthuis2021intersectiond}, and their relationship with the marginal extension problem is studied in \cite{bennequin2020extrafine}; a unifying perspective for Hilbert spaces can be found in \cite{sergeantperthuis2021interaction}. In \cite{sergeantperthuis2022regionalized}, we proposed a general framework for optimization of presheaves which in particular allows defining an Entropy functional for $\mathcal{A}$-specifications. In \cite{sergeantperthuis:hal-04456412}, we showed how the characterization of extreme Gibbs measures, one of the steps for proving a zero-one law for the extreme Gibbs measures, transfers to $\mathcal{A}$-specifications.

\section{Structure of the Paper and Contribution}

We start by recalling some important notions of rigorous statistical mechanics \cite{Georgii+2011} (specifications, Gibbs measures). We then show that the standard concept that encodes the statistical system, which is the notion of 'specifications', can be extended into a poset representation; we also show that the Gibbs measures of a statistical system are geometric invariants of this representation. From there, we propose a novel categorical formulation for statistical mechanics; we define `generalized specifications' and `generalized Gibbs measures' for those specifications: the $\mathcal{A}$-specifications. We characterize the generalized Gibbs measures for projective $\mathcal{A}$-specifications and their Gibbs measures. We then recall what extreme Gibbs measures are and their characterization on the tail $\sigma$-algebra. We extend such characterization to extreme Gibbs measures of $\mathcal{A}$-specifications. Finally, we recall what the Bethe free energy is in the context of graphical models and what the Belief Propagation algorithm is; when the graphical model is a Hidden Markov model (HMM), the Belief Propagation relates to Kalman filtering. We explain how we can define such variational free energy for $\mathcal{A}$-specifications. We then give a message-passing algorithm for finding critical points of the variational free energy of $\mathcal{A}$-specifications.

\section{Background: Rigorous Statistical Mechanics}

We will follow the presentation of Georgii's reference book \emph{Gibbs Measures and Phase Transitions} \cite{Georgii+2011}.

\begin{defn}[Markov Kernel]
A Markov kernel $k$ from the measurable space $(E,\mathcal{E})$ to the measurable space $(E_1,\mathcal{E}_1)$ is a function $k:\mathcal{E}_1\times E \to [0,1]$ such that 
\begin{enumerate}
    \item $\forall \omega\in E$, $k(.\vert \omega)$ is a measure on $E_1$
    \item $\forall A\in \mathcal{E}_1$, $k(A\vert .)$ is a measurable map from $E$ to $\mathbb{R}$
    \item $\forall \omega \in E$, $k(E_1\vert \omega)=1$, i.e., $k(.\vert \omega)$ is a probability measure.
\end{enumerate}
\end{defn}

We will denote a Markov kernel $k$ from $(E,\mathcal{E})$ to $(E_1,\mathcal{E}_1)$ as $k:(E,\mathcal{E})\to (E_1,\mathcal{E}_1)$. We denote $[(E,\mathcal{E}),(E_1,\mathcal{E}_1)]_K$, the set of kernels from $(E,\mathcal{E})$ to $(E_1,\mathcal{E}_1)$; if there is no ambiguity on the $\sigma$-algebras the spaces are provided with, we will simply denote it as $[E,E_1]_K$. We denote $\mathbb{P}(E)$, the space of probability distributions over $E$; it is a measurable space for the smallest $\sigma$-algebra that makes the evaluation maps, on measurable sets of $E$, measurable.\\

Markov kernels can be composed as follows. Let $k: E\to E_1$, $k_1:E_1\to E_2$ be two Markov kernels, then the composition $k_1\circ k:E\to E_2$ is the following Markov kernel: for any $A\in \mathcal{E}_2$ and $\omega\in E$,

\begin{equation}\label{chapitre-7-composition-kern}
k_1\circ k(A\vert \omega) =\int k_1(A\vert \omega_1) k(d\omega_1\vert \omega)
\end{equation}

 A measurable map $f:(E,\mathcal{E})\to (E_1,\mathcal{E}_1)$ between two measurable spaces can be extended into the following Markov kernel: for any $A\in \mathcal{E}_1$ and $\omega\in E$,

\begin{equation}
k_f(A|\omega)= 1[f(\omega) \in A]
\end{equation}

To avoid having too many notations, we will denote $k_f$ as $f$ and the context will specify if $f$ refers to the measurable map or its extension; for example, a composition $k\circ f$ between a Markov kernel $k$ and $f$ necessarily means that, here, $f$ refers to $k_f$. We will also denote $k_1\circ k$ as $k_1k$.

A probability measure $p\in \mathbb{P}(E)$ can also be identified with the following Markov kernel $k_p$ from $\ast$, the measurable space with one element, to $(E,\mathcal{E})$;  for any $A\in \mathcal{E}$, $k_p(A\vert \ast)= p(A)$. Similarly, we identify $p$ and $k_p$.
 
\begin{rem}
    Measurable spaces and measurable maps form a category. Giry \cite{Giry} and Lawvere \cite{Lawvere} are the first to have remarked that measurable spaces and Markov kernels also form a category; the latter category is the Kleisli category of the first. 
 \end{rem}

\begin{defn}[Proper Kernel, Section 1.1. \cite{Georgii+2011}]
Let $\mathcal{E}_1 \subseteq \mathcal{E}$ be two $\sigma$-algebras of a set $E$, a kernel $k\in [(E,\mathcal{E}_1),(E,\mathcal{E})]_K$ is proper if and only if, for any $A\in \mathcal{E}$, $B\in \mathcal{E}_1$,

\begin{equation}
k(A\cap B|.)=k(A|.)1_B
\end{equation}
\end{defn}

Let us set the notations. $I$ is the set of components of a complex system. $(E_i,\mathcal{E}_i,i\in I)$ is a collection of measurable spaces, with each $E_i$ being the space of configuration (state space) of the component $i\in I$. $(E,\mathcal{E})$ denotes the state space of the system; it is the product $E:=\prod_{i\in I} E_i$ with the product $\sigma$-algebra. For a sub-collection of components $a\subseteq I$, we denote $E_a:= \prod_{i\in a} E_i$ the associated state space, and $\mathcal{E}_a$ the associated product $\sigma$-algebra. $i^a:E\to E_a$ is the projection that sends a configuration $\omega:=(\omega_i,i\in I)$ to the configuration of the sub-collection $a$, $(\omega_i,i\in a)$. Finally, let us denote $\Pa_f(I)$ the set of finite subsets of $I$.

\begin{defn}[Specification, Adaptation of Def. 1.23 \cite{Georgii+2011}]\label{background:standard-specification}

A specification $\gamma$ with state space $(E,\mathcal{E})$ is a collection $(\gamma_a,a\in \Pa_f(I))$ of proper Markov kernels such that for any $a\in \Pa_f(I)$, $\gamma_a\in [(E_{\overline{a}},\mathcal{E}_{\overline{a}}), (E,\mathcal{E})]_K$ and which satisfies that for any $b\subseteq a$, i.e $\overline{a}\subseteq \overline{b}$ and any $A\in \mathcal{E}$,

\begin{equation}\label{compatibility-specification}
\gamma_b \circ i^{\overline{b}} \circ \gamma_{a}(A\vert .)=\gamma_a(A\vert .)
\end{equation}

\end{defn}

\begin{rem}
In Definition (1.23) \cite{Georgii+2011}, $E$ is the product over the same measurable space $X$ over $I$, and $I$ is a countably infinite set.
\end{rem}

In the standard definition of a specification (Definition \ref{background:standard-specification}), the Markov kernels encode border conditions for experiments that only involve finite numbers of components ($a\in \mathcal{P}_f(I)$). It is what formally encodes the statistical system. The next definition defines the ‘phases' of the system.

\begin{defn}[Gibbs measures, Def. 1.23 \cite{Georgii+2011}]
Let $\gamma$ be a specification with state space $E$; the set of probability measures,

\begin{equation}
\G(\gamma):=\{p \in \p(E): \quad \mathbb{E}_{p}[A|\mathcal{E}_{\overline{a}}](\omega)=\gamma_a(A|\omega_a) \ p \text{ a.s.}\}
\end{equation}

is the set of Gibbs measures of $\gamma$.
\end{defn}

One of the central problems of (rigorous) statistical mechanics is to understand the relationship between a specification $\gamma$ (statistical system) and its set of Gibbs measures $\G(\gamma)$ (its phases).

\section{Statistical Systems as poset representations}

\begin{thm}\label{poset-representation:thm}
Let $\gamma$ be a specification with state space $E$. For any $a,b\in \Pa_f(I)$ such that $b\subseteq a$, there is a unique Markov kernel $F^a_b:E_{\overline{a}}\to E_{\overline{b}}$ such that the following diagram commutes,

\begin{equation}\label{extension-specification-foncteur}
\begin{tikzpicture}[baseline=(current  bounding  box.center),node distance=2cm, auto]
\node (A) {$E_{\overline{a}}$ };
\node (B) [right of =A] {$E$};
\node (C) [below of=A] {$E_{\overline{b}}$};
\draw[->] (A) to node {$\gamma_a $} (B);
\draw[->] (A) to node [left]{$F^a_b$} (C);
\draw[->] (C) to node {$\gamma_b$} (B);
\end{tikzpicture}
\end{equation}

i.e. such that $\gamma_b\circ F^a_b=\gamma_a$. Furthermore for any collection $a,b,c\in \Pa_f(I)$ with $a\subseteq b\subseteq c$,

\begin{equation}
F^b_c\circ F^a_b= F^a_c
\end{equation}

\end{thm}

\begin{proof}

Let $b\subseteq a$, let $F^a_b$ satisfy the commutative diagram \ref{extension-specification-foncteur}, then,

\begin{equation}
i^{\overline{b}}\gamma_bF^a_b= i^{\overline{b}}\gamma_a
\end{equation}

therefore, 

\begin{equation}
F^a_b = i^{\overline{b}}\gamma_a
\end{equation}

For any $a,b\in \A$ such that $b\subseteq a$ let $F^a_b = i^{\overline{b}}\gamma_a$ then for $c\subseteq b\subseteq a$,

\begin{equation}
F^b_cF^a_b= i^{\overline{c}}\gamma_b i^{\overline{b}}\gamma_a
\end{equation}

Equation \ref{compatibility-specification} can be rewritten as, for $b\subseteq a$, 

\begin{equation}
\gamma_b i^{\overline{b}}\gamma_a= \gamma_a
\end{equation}

therefore,

\begin{equation}
F^b_cF^a_b= F^a_c
\end{equation}

\end{proof}

\begin{defn}[Partially ordered set]
A partially ordered set (poset), $(\A,\leq)$, is a set $\A$ provided with a binary relation $\leq: \A\times \A\to\{0,1\}$, such that
\begin{enumerate}
    \item reflexive: $\forall a\in \A, a\leq a$
    \item transitive: if $c\leq b$ and $b\leq a$ then $c\leq a$
    \item antisymmetric: if $b\leq a$ and $a\leq b$ then $a=b$
\end{enumerate}

\end{defn}

$(\Pa_f(I),\subseteq)$ is a poset for the inclusion relation; $(\Pa_f(I),\supseteq)$ is also a poset for the reversed inclusion relation: $b\supseteq a \iff a\subseteq b$. The convention is to denote $(\Pa_f(I),\supseteq)$ as $\Pa_f(I)^{op}$ because what relates $(\Pa_f(I),\supseteq)$ to $\Pa_f(I)$ is the fact that the order is ‘opposed'; the same convention holds for any poset; $\A^{op}$ is the set $\A$ with opposed order. \\

We call a representation of the poset $\A$: a collection of ‘spaces' $(G(a),a\in \A)$ and a collection of ‘maps' $(G^b_a:G(b)\to G(a); b,a\in \A, b\leq a)$ which satisfies for any $c\leq b\leq a$, $G^b_c\circ G^a_b= G^a_c$. We keep the notion of ‘poset representation' a bit vague for now (we do not say what ‘spaces' or ‘maps' are); we keep this notion vague at this stage but in the next section will make this notion formal by introducing the concept of category and of functor.

We call a representation of the poset $\mathcal{A}$: a collection of ‘spaces' $(G(a),a\in \mathcal{A})$ and a collection of ‘maps' $(G^b_a:G(b)\to G(a); b,a\in \mathcal{A}, b\leq a)$ which satisfies for any $c\leq b\leq a$, $G^b_c\circ G^a_b= G^a_c$. We keep the notion of ‘poset representation' a bit vague for now (we do not say what ‘spaces' or ‘maps' are); we keep this notion vague at this stage but in the next section will make this notion formal by introducing the concept of category and of functor.

Theorem \ref{poset-representation:thm} implies that a specification $\gamma$ can be promoted to a representation $F$ of $(\Pa_f(I),\supseteq)$.

\begin{thm}\label{poset-represnetation:limit}
Let $\gamma$ be a specification with state space $E$, let $p\in \G(\gamma)$. For any $a\in \A$ let $p_a:=  i^{\overline{a}}\circ p$; it is the marginal distribution on $E_{\overline{a}}$ of $p$. Then, for any $a,b\in \Pa_f(I)$ such that $b\subseteq a$,

\begin{equation}
F^a_b\circ p_a= p_b
\end{equation}

\end{thm}

 \begin{proof}
For any $a,b\in \Pa_f(I)$ such that $b\subseteq a$,

\begin{equation}
\gamma_b\circ( F^a_bp_a )= p
\end{equation}

therefore, $F^a_bp_a= i^{\overline{b}}p$ and 

\begin{equation}
F^a_b p_a= p_b
\end{equation}
   
\end{proof}

\section{Categorical formulation of specification and Gibbs measures}

We denote categories in bold, e.g. $\mathbf{C}$.
We will denote $\mathbf{Mes}$ as the category that has as objects measurable spaces and as morphisms measurable maps (Section 1 \cite{Giry}); we will denote $\mathbf{Kern}$ as the category that has as objects measurable spaces and as morphisms Markov kernels (the Kleisli category of the monad $\mathbf{Mes}$).

A poset, $(\mathcal{A},\leq)$, can be seen as a category, $\mathbf{A}$, with at most one morphism between two objects: the objects of $\mathbf{A}$ are the elements of $\mathcal{A}$ and for any two elements $b,a\in \mathcal{A}$ there is one morphism $b\to a$ when $b\leq a$. From now on, we will drop the bold notation for the category $\mathbf{A}$ and denote it simply as $\mathcal{A}$. A functor $G$ from a poset $\mathcal{A}$ to a category $\mathbf{C}$ is precisely a collection of maps $G^b_a$ for $b,a\in \mathcal{A}$ such that $b\leq a$, which satisfy $G^b_a\circ G^c_b = G^c_a$ for any three elements $c\leq b\leq a$. A functor from $\mathcal{A}$ to some target category is what we will call a representation of the poset $\mathcal{A}$; in general, the target category is the category of vector spaces or modules \cite{Simson1993LinearRO}. For this article, the target category will be $\mathbf{Mes}$ and $\mathbf{Kern}$.

Consider a functor $G:\A\to \mathbf{Set}$ from a poset $\A$ to the category of sets. A collection $(\omega_a,a\in \A)$ is called a section of $G$ if for any $b\leq a$, $G^b_a(\omega_b)=\omega_a$; the set of sections of a poset representation is called the limit of $G$ ( III.4 \cite{mac2013categories}) and denoted $\lim G$. It is an ‘invariant' of $G$ that can be computed using homological algebra when the target category of $G$ is enriched with some algebraic structure (Chapter 13 \cite{gallier2022homology}).

In Theorem \ref{poset-representation:thm} we showed that we can associate to a specification $\gamma$, a functor from $(\Pa_f(I),\supseteq)$ to $\Kern$. The convention is to call a functor with source $\A^{op}$, a \emph{presheaf}. In Theorem \ref{poset-represnetation:limit} we showed that Gibbs measures of $\gamma$ are ‘sections' of $F$. We will denote this set  of ‘sections' as $[\ast, F]_{K,\A}$; more precisely for a functor $F:\A\to \Kern$,

\begin{equation}\label{mes-sections}
[\ast, F]_{K,\A} := \{ (p_a \in \p(F(a)), a\in \A) \vert \quad \forall b\leq a, F^b_ap_b= p_a\} 
\end{equation}

Introducing the presheaf $F$ and $[\ast, F]_{K,\A}$ is our way to emphasize that the compatible measures $p\in [\ast, F]_{K,\A}$ don't have to be measures over the whole space $E$. There is no need to define statistical systems ‘globally', one can also define them ‘locally'.

Let us now introduce the more general, categorical setting we propose for statistical systems.

\begin{defn}[Generalized Specification, $\A$-Specifications]
Let $\A$ be a poset, a generalized specification over $\A$, or simply $\A$-specification, is a couple $(G,F)$ of a presheaf and a functor where $G:\A^{op}\to \mes$ and $F:\A\to \Kern$ are such that for any $a,b\in \A$ with $b\leq a$, 

\begin{equation}
G^{a}_{b} F^{b}_{a}=\id
\end{equation}
\end{defn}

In the previous definition, $G$, in the particular case of $E=\prod_{i\in I}E_i$, encodes the collection of projections $i^a_b:E_a\to E_b$ for $b\subseteq a$; it is in some way the ‘skeleton' of the spaces of observables of the statistical system. It is a key ingredient for the generalization of (rigorous) statistical mechanics to a categorical framework.

\begin{defn}[Gibbs measures for $\A$-specifications]
Let $\gamma=(G,F)$ be an $\A$-specification, we call the Gibbs measures of $\gamma$ the sections of $F$,

\begin{equation}
\G_g(\gamma):= [\ast, F]_{K,\A}
\end{equation}

\end{defn}


\section{Gibbs measure of projective $\A$-specifications}

Let $E$ be a measurable space; we denote $L^\infty(E)$ the set of bounded, real-valued, measurable functions over $E$.
One associates to a Markov kernel $F:E_b\to E_a$ a linear map $\pi:L^{\infty}(E_a)\to L^{\infty}(E_b)$ defined as follows: for any $f\in L^{\infty}(E_a)$,

\begin{equation}
\forall \omega_b\in E_b,\quad \pi(f)(\omega_b)= \int f(\omega_a) F(d\omega_a\vert  \omega_b)
\end{equation}

This association is `functorial', we may denote the underlying functor $L^{\infty}: \mathbf{Kern}^{op}\to \mathbf{Vect}$ which is presheaf from the category of Markov kernels to the category of vector spaces. It is the presheaf that associates spaces to their space of observables. Let us denote $L^{\infty}\circ G: \A\to \mathbf{Vect}$ as $i$ and $L^{\infty}\circ F: \A^{op}\to \mathbf{Vect}$ as $\pi$. In these notation one has that for any $a,b\in \A$ such that $b\leq a$ then $\pi^a_b\circ i^b_a=\id$.

For the definition and characterization of projective presheaf over a poset see \cite{YANAGAWA2001341,https://doi.org/10.48550/arxiv.2112.02609,https://doi.org/10.48550/arxiv.2010.09651}.

\begin{defn}[Projective $\A$-specifications]
An $\A$-specification $(G,F)$ is called projective when $L^\infty\circ F$ is a projective presheaf (in $\textbf{Vect}$). In other words, there is a collection of presheaves $(S_a,a\in \A)$ such that, $L^\infty\circ F\cong \bigoplus_{a\in \A} S_a$ where for any $b\geq a$, $S_a(b)$ is a constant vector space denoted $S_a$ and ${S_a}^b_c =\id $ for any $a\leq c\leq b$ and $S_a(b)=0$ if  $b\not \geq a$. The collection of presheaves $(S_a,a\in\A)$ is called the decomposition of $(G,F)$.
\end{defn}

For any poset $\A$, symmetrizing the order defines the following equivalence relation,

\begin{equation}
\forall a,b\in \A, \ a\sim b \iff a\leq b \ \text{or} \ b\leq a
\end{equation}

The equivalence classes of this equivalence relation are the connected components of $\A$ that we will denote as $\C(\A)$.  To each element of $a\in \A$ one can associate its connected component $\C(a)$. If each connected component has a minimum element, in other words, if for any $C\in \C(\A)$, and any $b\in C$, there is $c\in C$ such that, $c\leq b$, then we shall denote, $\C_\ast(\A)$ as the collection of these minimum elements; if not $\C_\ast(\A)= \emptyset$.

To conclude this article let us characterize Gibbs measures of projective $\A$-specifications.\\

\begin{thm}\label{chapitre-7:decomposability-gibbs-measures}
Let $\gamma=(G,F)$ be a projective $\A$-specification. If at least one of the connected components of $\A$ does not have a minimum element, i.e. when,

\begin{equation}
\C_\ast(\A)=\emptyset
\end{equation}

then,

\begin{equation}
\G_g(\gamma)=\emptyset
\end{equation}

if not,
\begin{equation}
\G_g(\gamma)= \prod_{a\in \C_\ast(\A)} \Pro(G(a))
\end{equation}

\end{thm}

\begin{proof}

Let us denote $L^\infty G$ as $i$ and, $L^\infty F$ as $\pi$; $(i, \pi)$ is decomposable, let $(S_a,a\in \A)$ be its decomposition. For any $a,b\in \A$ such that $b\leq a$ and $\mu\in \G(\gamma)$, let us denote $L^\infty\mu$ as $\nu$. For any $v\in L^\infty F(a)$,

\begin{equation}
\nu_b\pi^a_b(\sum_{c\leq a} S_c(a)(v))= \nu_a(\sum_{c\leq a} S_c(a)(v))
\end{equation}

where $S_c(a)(v)$ denotes the projection of $v$ on $S_c(a)$; therefore,

\begin{equation}
\nu_b(\sum_{c\leq b} {S_c}^a_b(v))=\nu_a i_a^b(\sum_{c\leq b} {S_c}^a_b(v))=  \nu_a \left(\sum_{c\leq a} S_c(a)(v)\right)
\end{equation}

and so,

\begin{equation}
\nu_a(\sum_{\substack{c\leq a \\ c\not \leq b}} S_c(a))=0
\end{equation}

Therefore for any $a\not \in \C_\ast(\A)$, $\nu_a|_{S_a(a)}=0$. Furthermore, 

\begin{equation}
\colim i \cong \bigoplus_{a\in \A} S_a(a).
\end{equation}

$\nu$ is uniquely determined by $(\nu_a|_{S_a(a)},a\in \A)$; if there is a connected component $C\in \C(\A)$ that does not have a minimal element, for any $a\in C$, 

\begin{equation}
\nu_a|_{S_a}=0
\end{equation}

Therefore for any $a\in C$, $\nu_a=0$; this is contradictory with the fact that $\mu_a\in \Pro(\gamma(a))$ and so,

\begin{equation}
\G_g(\gamma)=\emptyset
\end{equation}

When $\C_\ast(\A)$ is non empty for any functor, $H$, from $\A$ to $\Set$, 

\begin{equation}
\lim H\cong \prod_{a\in \C_\ast(\A)} H(a)
\end{equation}

therefore,

\begin{equation}
\G_g(\gamma)= \prod_{a\in \C_\ast(\A)} \Pro(G(a))
\end{equation}

\end{proof}

\section{Characterization of extreme Gibbs measures of $\A$-specifications}

We now turn to the characterization of extreme Gibbs measures (Theorem 7.7 \cite{Georgii+2011}), which is one of the steps for proving a zero-one law for the extreme Gibbs measures. We will show how it transfers to $\A$-specifications. In the classical theory of rigorous statistical mechanics, the tail $\sigma$-algebra generates the observables for which a `generalized' law of large numbers (zero–one law) holds. Following \cite{sergeantperthuis:hal-04456412}, we give a candidate for such a $\sigma$-algebra in the categorical setting and show the associated extreme Gibbs measures decomposition.

Let us recall the definition of the tail $\sigma$-algebra in the classical formulation. Let us first consider the case of time series, i.e. $\Omega= \prod_{i\in \mathbb{N}}E_i$ with $E_i$ are measurable spaces; let us denote $\mathcal{E}_{\geq k}$ as the $\sigma$-algebra generated by the cylinders $\prod_{n\geq k} E_k$. The tail $\sigma$-algebra is defined as $\bigcap_{k\in \mathbb{N}} \mathcal{E}_{\geq k}$. For specifications, $I$ is any set and $\mathcal{E}_{\overline{a}}$ is indexed by a subset $a\subseteq I$ that is finite. We defined $\mathcal{E}_{\overline{a}}$ to be a $\sigma$-algebra of $E_{\overline{a}}$, however it can also be identified with the smallest $\sigma$-algebra on $E$ that make $i^{\overline{a}}:E, \mathcal{E}\to E_{\overline{a}},\mathcal{E}_{\overline{a}}$ measurable. Through this identification one defined $\mathcal{E}_{\infty}:= \cap_{a\in \mathcal{P}_f(I)}\mathcal{E}_{\overline{a}}$.\\

For a functor from $\mathcal{A}$ to $\mathbf{Mes}$, let us denote $\sigma(G(a))$ as the $\sigma$-algebra of the measurable space $G(a)$, where $a\in \mathcal{A}$, and $\sigma(G)$ as the underlying functor defined as $\sigma(G)^b_a A_b:= {G^a_b}^{-1} A_b$, with $b\leq a$. We propose that one candidate that plays the role of the tail $\sigma$-algebra for a given specification $\gamma=(G,F)$ is $\lim \sigma(G)$ defined as, 

\begin{equation}
\lim \sigma (G):= \{(A_a\in \sigma(G(a)),a\in \mathcal{A})\vert \forall a,b\in \mathcal{A}, \quad A_a= {G^a_b}^{-1} A_b\}
\end{equation}

Let us denote \(1_A: E \to \{0,1\}\) the indicator function over the set \(A\) that sends \(\omega \in A\) to \(1\) and \(\omega \not\in A\) to \(0\). Let us remark that \(1_{A_b} \circ G^a_b = 1[G^a_b(\omega_a) \in A_b] = 1_{{G^a_b}^{-1} A_b}\). Remark that \(A \in \lim \sigma (G)\) is equivalent to \(1_A \in \lim i\); in other words, \(\lim \sigma (G)\) is the restriction of \(\lim i\) to indicator functions of the form \(1_{A_a}, a \in \mathcal{A}\).

Finally, we also need to recall that for any $f\in L^{\infty}(E)$ and $\mu \in \mathbb{P}(E)$, one can define a measure $f \cdot \mu$ as $f \cdot \mu(d\omega)= f(\omega)d\omega$.\\

The key proposition of this section is Proposition \ref{prop:key-result}; the proof of that proposition is given for $G(a), a\in \mathcal{A}$ finite measurable sets. \\

\textbf{Assumption:} Therefore, we assume in what follows that the measurable sets $G(a)$ are finite. \\

However, there is no finiteness constraint on $\mathcal{A}$. A weaker version holds when $G(a)$ is not finite. We will say that $F > 0$ when for any $a, b \in \mathcal{A}$, such that $b \leq a$, $F(\omega_a \vert \omega_b) > 0$ for any $\omega_b$ such that $G^a_b(\omega_a)=\omega_b$; $G\circ F=\id$ requires that $F(\omega_a \vert \omega_b)=0$ when $G^a_b(\omega_a)\neq\omega_b$.\\

The following lemma is an extension of the classical result that states that conditioning over a $\sigma$-subalgebra $\mathcal{F}_1 \subseteq \mathcal{F}$ defines a morphism of modules when finer ($\mathcal{F}$ measurable) observables are seen as modules over the coarser ($\mathcal{F}_1$ measurable observables).

\begin{lem}\label{lem:ring-morph}
Let $E_1, E_2$ be two measurable spaces, let $g:E_2\to E_1$ be a measurable map and $f:E_1\to E_2$ be a Markov kernel so that, $f\circ g=\id$. Let us denote respectively $i$ and $\pi$ the induced linear maps on $L^{\infty}(E_1),L^{\infty}(E_2)$. Let $h\in L^{\infty}(E_2)$ and $k\in  L^{\infty}(E_1)$, then,

\begin{equation}
\pi(h).k= \pi(h. i(k))
\end{equation}
\end{lem}

\begin{proof}

Let us first prove the result in the particular case when $h= 1_B$ with $B\in \sigma(E_2)$ and $k=1_A$ with $A\in \sigma(E_1)$. Let us denote $\overline{A}$ the complement of $A$, then $1_A+1_{\overline{A}}=1$ and $1_A. 1_{\overline{A}}=0$. Furthermore $i(1_A)= 1_{g^{-1}A}$

\begin{equation}
\pi(1_B)= \pi(1_B.1_{g^{-1}A})+\pi(1_B.1_{\overline{g^{-1}A}})
\end{equation}

and

 \begin{align}
\pi(1_B.1_{g^{-1}A})\leq & \pi(1_{g^{-1}A})= \pi\circ i(1_A)= 1_A \\
\pi(1_B.1_{\overline{g^{-1}A}})\leq &\pi(1_{\overline{g^{-1}A}})= 1_{\overline{A}}
 \end{align}

Therefore, 
\begin{equation}
\pi(1_B)1_A= \pi(1_B.1_{g^{-1}A})1_A+\pi(1_B.1_{g^{-1}A})1_A
\end{equation}

But, $\pi(1_B.1_{g^{-1}A})1_A\leq 1_{\overline{A}}.1_A=0$ so, $\pi(1_B)1_A=\pi(1_B.1_{g^{-1}A})1_A$. Furthermore, $\pi(1_B.1_{g^{-1}A})= \pi(1_B.1_{g^{-1}A})1_A + \pi(1_B.1_{g^{-1}A})1_{\overline{A}}$ therefore, 

\begin{equation}
\pi(1_B.1_{g^{-1}A})1_{\overline{A}}\leq \pi(i(1_A)).1_{\overline{A}}=0
\end{equation}

We just showed that, 

\begin{equation}
\pi(1_B.1_{g^{-1}A})=\pi(1_B.1_{g^{-1}A})1_{\overline{A}}
\end{equation}

So $\pi(1_B).1_A= \pi(1_B.i(1_A))$. The result then extends by linearity directly to $h=\sum_{k\leq n} 1_{B_k}$ and $k= \sum_{k\leq n_1} 1_{A_n}$, which ends the proof.

\end{proof}

Let us remark that if $A\in \lim \sigma(G)$ then $\overline{A}:=(\overline{A}_a, a\in \A)$ is also in $\lim \sigma(G)$.

\begin{prop}\label{prop:key-result}
Let $\gamma=(G,F)$ be a specification, let $G(a)$ be finite sets for any $a\in \A$, let $F>0$. Let $\mu \in \mathcal{G}(\gamma)$, for any $f\in \prod_{a\in A}L^{\infty}(G(a))$,  such that $\forall a\in \A, \mu_a(f_a)=1$,

\begin{equation}
f.\mu \in \mathcal{G}_g(\gamma)\iff \exists \tilde{f} \in \lim i, \text{ s.t. } f.\mu= \tilde{f} .\mu
\end{equation}

\end{prop}

\begin{proof}

Let us assume that $f.\mu \in \mathcal{G}_g(\gamma)$, then for any $a, b \in \mathcal{A}$ such that $b \leq a$, and for any $g_a \in L^{\infty}(G(a))$, by hypothesis, $(f.\mu)_b \pi^a_b(g_a) = (f.\mu)_a (g_a)$; it can be rewritten as,

\begin{equation}
\mu_b\left(\pi^a_b(g_a) . f_b\right) = \mu_a\left(f_a . g_a\right)
\end{equation}

By Lemma \ref{lem:ring-morph}, $\pi^a_b(g_a) . f_b = \pi^a_b(g_a . i^b_a f_b)$; therefore, 
\begin{equation}
\mu_b \pi^a_b(g_a . i^b_a f_b) = \mu_a\left(f_a . g_a\right)
\end{equation}

Therefore $f_a= i^b_af_b$ $\mu_a$-almost surely.

We will now show that there is $\tilde{f}\in \lim i$ such that $\tilde{f}.\mu = f.\mu$. It is in this part of the proof that we assume that $G(a)$ are finite sets and that $F > 0$. Let's call $S_a := \text{supp} \mu_a$, the support of $\mu_a$, i.e., the set $\text{supp} \mu_a := \{\omega_a\in G(a) \vert \mu_a(\omega_a) > 0\}$. Let us denote $N_a := \overline{S_a}$ its complement and $M_a = 1_{N_a}$. We will now show that $(N_a, a \in \mathcal{A}) \in \lim \sigma(G)$.

For any $b, a \in \mathcal{A}$ such that $b \leq a$, $\mu_a i^b_a = \mu_b$; therefore, as $\mu_b(M_b) = 0$, one has that $\mu_a i^b_a(M_b) = 0$. Recall that $i^b_a(M_b)$ is the indicator function of the set ${G^a_b}^{-1}N_b$; the previous remark implies that $i^b_a(M_b) \leq M_a$.
Furthermore, $\mu_b \pi^a_b (M_a) = \mu_a(M_a) = 0$; therefore, $\pi^a_b(M_a) \leq M_b$.

Hence, as $\pi^a_b(i^b_a M_b) = M_b$ and $i^b_a M_b \leq M_a$, then by applying $\pi^a_b$ on both sides, $M_b \leq \pi^a_b(M_a)$. And so $\pi^a_b(M_a) = M_b$.

Recall that we showed that $\pi^a_b(M_a) = M_b$ and $i^b_a(M_b) \leq M_a$. In particular, $M_a - i^b_a(M_b) \geq 0$; furthermore, $\pi^a_b(M_a - i^b_a(M_b)) = 0$ so $M_a = i^b_a(M_b)$. To be more explicit: $\forall \omega_b\in G(b)$,

\begin{equation}
\pi^a_b(M_a - i^b_a(M_b))(\omega_b) = \sum_{\omega_a \in G(a)} F(\omega_a \vert \omega_b) [M_a - i^b_a(M_b)](\omega_a)
\end{equation}

As, by hypothesis, for any $\omega_b$ such that $G^a_b(\omega_a)=\omega_b$ one has that $F(\omega_a \vert \omega_b) > 0$, then $M_a = i^b_a(M_b)$. This implies that $M \in \lim i$ and $N \in \lim \sigma(G)$. This also implies that $S \in \lim \sigma(G)$. 

Let $\tilde{f} = f1_{S}$. Then $f1_S \in \lim F$ and for any $a \in \mathcal{A}$, $f_a = \tilde{f}_a$ $\mu_a$-a.s., which ends the proof.

\end{proof}

One remarks that $\lim i$ is a subset of $\lim \pi$: for any $f\in \lim i$, by definition for any $a,b\in \A$ such that $b\leq a$, $i^b_a f_b=  f_a$ and so $\pi^a_bi^b_a f_b= \pi^a_b f_a$ so $f_b=\pi^a_b f_a$.

Let us also remark that for any $b\leq a\in \A$, $\mu_a(A_a)=\mu_b(A_b)$ for $\mu \in \mathcal{G}(\gamma)$, $A\in \lim G$.

\begin{thm}[Extreme measure characterisation (Generalisation of Theorem 7.7 \cite{Georgii+2011}) ]
Let $\gamma=(G,F)$ be a specification, let $G(a)$ be finite sets for any $a\in \A$, let $F>0$. $\mathcal{G}_g(\gamma)$ is a convex set. Each $\mu\in \mathcal{G}_g(\gamma)$ is uniquely determined by it's restriction to $\lim \sigma(G)$. Furthermore $\mu$ is extreme in $\mathcal{G}(\gamma)$ if and only if for any $A\in \lim \sigma(G)$, $\forall a\in \A$, $\mu_a(A_a)=0$ or $1$.

\end{thm}

\begin{proof}
Let us denote $\pi^{*}$ the functor from $\A$ to $\Vect$ for which for any $b\leq a$, ${\pi^{*}}^b_a: (L^{\infty}F(b))^{*}\to (L^{\infty}F(a))^{*} $ is the dual of $\pi^a_b$ that send linear forms to linear forms. Then $\mathcal{G}_g(\gamma)$ is a subspace of the vector space $\lim F^{*}$ and furthermore for any $a\in \A$ and $p\in [0,1]$, $p\mu_a+ (1-p) \nu_a\in \mathbb{P}(G(a))$ whenever $\mu_a,\nu_a\in \mathbb{P}(G(a))$. Therefore $\mathcal{G}_g(\gamma)$ is a convex set. 

Proposition \ref{prop:key-result}, allows us to apply a similar proof, when done with caution, to the one found of Theorem 7.7 in \cite{Georgii+2011}. Let us recall the proof. Let $\mu,\nu\in \mathbb{G}(\gamma)$ such that $\mu_{\vert \lim i}= \nu_{\vert \lim i}$. Let $\overline{\mu}= \frac{\mu+\nu}{2}$, then $\overline{\mu}\in \mathcal{G}(\gamma)$. But $\mu$ and $\nu$ are absolutely continuous with respect to $\overline{\mu}$ therefore for any $a\in \A$ there is $f_a,g_a\in L^{\infty}(G(a))$ such that $\mu_a= f_a \overline{\mu}_a$ and $\nu_a= g_a \overline{\mu}_a$. By Proposition \ref{prop:key-result}, $f,g\in \lim i$. By hypothesis, for any $h\in \lim i$ $\overline{\mu}_a(h_a)= \mu_a(h_a)=\nu_a(h_a)$. Importantly $i^b_a$ is a ring morphism of $L^{\infty}(G(b))$, i.e. $i^b_a(k_b.h_b)=i^b_a(k_b).i^b_a(h_b)$. Therefore for any $h,k\in \lim i$, $k.h\in \lim i$; as $f-g\in \lim i$, then it is also the case that $(f-g)^2\in \lim i $; but for any $a\in \A$, 

\begin{equation}
\overline{\mu}_a[(f_a-g_a)^2]=0
\end{equation}

so $f_a=g_a\quad \overline{\mu}_a-\text{a.s}$. Therefore $f\overline{\mu}= g\overline{\mu}$ and $\mu= \nu$.

Showing that $\mu\in \mathcal{G}_g(\gamma)$ is extreme is equivalent to $\mu$ being trivial on $\lim i$ is a direct generalization of Corollary 7.4 \cite{Georgii+2011} thanks to Proposition \ref{prop:key-result}. Let $\mu\in \mathcal{G}(\gamma)$ be not trivial on $\lim \sigma(G)$ then there is $A=(A_a,a\in\A)\in \lim \sigma(G)$ such that,

\begin{equation}
\forall a\in \A,\quad 0<\mu_a(A_a)<1 
\end{equation}

Therefore for any $a\in \A$, 

\begin{equation}
\mu_a= \mu_a(A_a)\frac{1_{A_a}.\mu_a}{\mu_a(A_a)} +\mu_a(\overline{A}_a)\frac{1_{\overline{A}_a}.\mu_a}{{\mu_a(\overline{A}_a)}}
\end{equation}

Furthermore, for any $b\leq a$,

\begin{equation}
i^b_a\frac{1_{A_b}}{\mu_a(A_b)}= \frac{1_{A_a}}{\mu_a(A_b)}= \frac{1_{A_a}}{\mu_a(A_a)}
\end{equation}

Therefore $\frac{1_{A_a}}{\mu_a(A_a)}, a\in \A$ is in $\lim i$ and so by Lemma \ref{lem:ring-morph}, $\left(\frac{1_{A_a}.\mu_a}{\mu_a(A_a)}, a\in \A\right) \in \mathcal{G}_g(\gamma)$. Similarly $\left(\frac{1_{\overline{A}_a}.\mu_a}{\mu_a(\overline{A}_a)}, a\in \A\right) \in \mathcal{G}_g(\gamma)$. In particular there is $0<p<1$ so that $\mu= p\nu +(1-p)\nu_1$ with $\nu, \nu_1\in \mathcal{G}_g(\gamma)$. Therefore $\mu$ is not an extreme measure. 

Assume now that $\mu \in \mathcal{G}_g(\gamma)$ is such that for any $A \in \lim \sigma(A)$, and any $a \in \mathcal{A}$, $\mu_a(A_a) = 0$ or $1$. Suppose that there is $0 < p < 1$ such that $\mu = p\nu + (1-p)\nu_1$ with $\nu, \nu_1 \in \mathcal{G}_g(\gamma)$. Then for any $a \in \mathcal{A}$, $\nu_a, {\nu_1}_a$ is absolutely continuous with respect to $\mu_a$. Therefore, there are $(f_a \geq 0, a \in \mathcal{A})$, $(g_a \geq 0, a \in \mathcal{A})$ both in $\prod_{a \in \mathcal{A}} L^{\infty}(G(a))$ such that $\nu = f\mu$ and $\nu_1 = g\mu$. As $\nu, \nu_1 \in \mathcal{G}_g(\gamma)$, then by Lemma \ref{lem:ring-morph}, $f, g \in \lim i$. Therefore, for all $a \in \mathcal{A}$, $\mu_a(f_a) = 0$ or for all $a \in \mathcal{A}$, $\mu_a(g_a) = 0$. So, $\mu = \nu$ or $\mu = \nu_1$ and $\mu$ is extreme in $\mathcal{G}_g(\gamma)$.

\end{proof}

Let us remark that if $\mathcal{A}$ has only one connected component, then for $A \in \lim \sigma(G)$, satisfying $\forall a \in \mathcal{A}, \mu_a(A_a) = 0$ or $1$ is equivalent to $\exists a \in \mathcal{A}, \mu_a(A_a) = 0$ or $1$. Indeed, if $a, b$ are in the same connected component, i.e., $a \leq b$ or $b \leq a$, then $\mu_a(A_a) = \mu_b(A_b)$.

\section{Background: Variational inference for graphical model and more}

Consider a joint distribution $P_{X,Y}\in \mathbb{P}(E\times E_1)$ over two random variables $X\in E, Y\in E_1$. A classical problem is, given an observation $\omega_1$ on $Y$, to compute the posterior $P_{X\vert Y}(\omega,\omega_1)= \frac{P_{X,Y}(\omega,\omega_1)}{P_Y(\omega_1)}$ with $P_Y(\omega_1)=\sum_{\omega\in E}P_{X\vert Y}(\omega,\omega_1)$, the marginal distribution of $Y$. However, doing so requires summing over all possible configurations of $X$, which can be computationally too costly. This is the case, for example, when $X=X_0,...,X_T$ with $X_i\in S$ and $E=\prod_{i\leq T}B$. Instead, one resorts to variational inference to compute $P_{X\vert Y}$ approximately \cite{alquier_approximate_2020}. We will now explain what variational inference is, but first let us introduce entropy and Gibbs free energy. When $E$ is a finite set, the entropy of a probability distribution $Q$ on $E$ is defined as:

\begin{equation}
S(Q)=-\sum_{\omega \in E} Q(x)\ln Q(x)
\end{equation}

Let $H$ be a measurable function $H:E\to \R$. For $Q\in \mathbb{P}(E)$, one calls $\mathbb{E}_{Q}[ H] - \frac{1}{\beta}S(Q)$ the Gibbs free energy; in general $\beta=1$. An important property is that, 

\begin{equation}\label{Maxent}
-\ln \sum_{\omega\in E} e^{-\beta H(\omega)}= \inf_{Q\in \p(E)} \mathbb{E}_{Q}[\beta H] - S(Q) 
\end{equation}

The optimal solution to Equation \ref{Maxent} is given by the Boltzmann distribution 
\begin{equation}
Q^{*}(\omega)= \frac{e^{-H(\omega)}}{\sum_{\omega\in E} e^{-\beta H(\omega)}}
\end{equation}

Let $H(\omega)=-\ln P_{X,Y}(\omega,\omega_1)$ and $\beta=1$, then $Q^{*}(\omega)= P_{X\vert Y}(\omega\vert \omega_1)$. Therefore, solving the optimization problem of Equation \ref{Maxent} is equivalent to computing the posterior $P_{X\vert Y}(\omega\vert \omega_1)$. Solving Equation \ref{Maxent} over a subset of distributions $Q\in \Theta\subseteq \mathbb{P}(E)$ is called variational inference. If furthermore the Gibbs free energy is replaced by an approximation, we call it approximate variational inference.

One remarks that $\inf_{Q\in \p(E)} \mathbb{E}_{Q}[\beta H] - S(Q)$ is equivalent to $\sup_{Q\in \p(E)}  S(Q)-\mathbb{E}_{Q}[\beta H]$. And this last optimization problem relates, through Lagrange multipliers, to maximizing entropy under energy constraints $U\in \R$,

\begin{equation}\label{Maxent-2}
\sup_{\substack{Q\in \mathbb{P}(E)\\ \mathbb{E}_Q [H]=U}} -\sum_{x\in E} Q(x)\ln Q(x)
\end{equation}

In the physics literature, one refers to Equation \ref{Maxent-2} as MaxEnt \cite{Kesavan2009}, which stands for the principle of maximum entropy and such principle has many application see \cite{Kesavan2009,DEMARTINO2018e00596}. Variational inference is called the variational principle.

Graphical models translate relations on graphs into conditional independence relations between variables \cite{10.5555/534975,lauritzen1996graphical,sergeantperthuis2021bayesiangraphoid}. The Hammersley-Clifford Theorem states that for strictly positive distributions, these conditional independence relations are equivalent to factorizing the joint distributions on the cliques of the graph. When a joint distribution factors according to an acyclic graph, computing the posterior $P_{X\vert Y}$ over some nodes of the graph can be done efficiently through dynamic programming with an algorithm called Belief Propagation. In the particular case where the graphical model is a Gaussian Hidden Markov Model, this dynamic programming algorithm is the smoother Kalman filter \cite{lecture2013probabilistic}.

The Belief Propagation algorithm solves an approximate variational inference problem for a variational free energy called the Bethe free energy \cite{doi:10.1143/JPSJ.12.753,Mezard,PeltrePhD,sergeantperthuis2022regionalized}. 

We now specify the previous statement and explain what we mean by a probability distribution that factors according to a collection of subsets of variables. Then, we recall the remarkable property that for probability distributions that factor according to an acyclic graph, entropy can be decomposed into sums of 'local' entropies. This will allow us to introduce what the Bethe free energy is and how it relates to entropy.

It is important to consider the case of graphical models to understand the more general setting of \emph{factor graphs} or \emph{factorization models} Yedidia, Freeman, Weiss consider in their seminal article, \emph{Constructing Free Energy Approximations and Generalized Belief Propagation Algorithms}, extending the correspondence between a Generalized Belief Propagation algorithm and the associated variational free energy that we will call the \emph{Generalized Bethe free energy}. We present their work and O. Peltre's complemented version and extension of their results (see SYCO 8 talk, \cite{PeltrePhD,10.1007/978-3-030-26980-7_23}) in the next section. Doing so will then motivate the entropy we introduce for $\mathcal{A}$-specifications and give intuition on the message passing algorithm we introduce for such specifications.

Let $E=\prod_{i\in I}E_i$ where $I$ is finite. $\mathbb{P}$ is said to factor according to a subset $\A$ of $\mathcal{P}(I)$ when for any $a\in \A$ there is $f_a:E_a\to \mathbb{R}$ such that, 
\begin{equation}
P(\omega)= \prod_{a\in \A} f_a(\omega_a)
\end{equation}

We will denote the space of probability distributions that factor according to $\A$ as $\Fac_{\A}$. We will call such space the space of ($\A$-)factorization models.

A graph $G=(I,A)$ is a collection of vertices $I$ and edges $A$; one can see it as a poset $\A(G)$ with the relation of inclusion. In other words, for any $v\in I$ and $e\in A$, $v\leq e$ if and only if $v\in e$.

Let us now recall a proposition that states that a strictly positive probability distribution on a finite set $E$, i.e., $Q(\omega)>0$ for $\omega\in E$, factors according to its marginal distribution on the edges and vertices \cite{Speed,yedidia_constructing_2005}. We denote $\p_{>0}$ as the space of strictly positive probability distributions.

\begin{prop}[Factorization on acyclic graphs]\label{factorisation-marginalisation-on-DAG}
Let $I$ be a finite set and let $E=\prod_{i\in I}E_i$ be a product of finite sets, and let $G=(I,A)$ be a finite acyclic graph. $Q\in \p_{>0}(E)$ factors accordingly to $\A(G)$, i.e., $Q\in G_{\A(G)}$ if and only if for any $\omega\in E$,
\begin{equation}
Q(\omega) = \frac{\prod_{e\in A}{i^e}_{*}Q(\omega_e) }{\prod_{v\in I} {i^v}_{*}Q^{d(v)-1}(\omega_v)},
\end{equation}
where $i^e:E\to E_e$ is the projection onto the state space $E_{v_1}\times E_{v_2}$ associated with the edge $e=(v_1,v_2)$, and similarly for $i^v:E\to E_v$.
\end{prop}

For a given sub-poset $\mathcal{A} \subseteq \mathcal{P}(I)$, let us denote
\begin{equation}
[\ast, i]_{\mathcal{A}} := \{ (Q_a \in \mathcal{P}(E_a), a \in \mathcal{A}) \, \vert \, \forall b \leq a, \, i^b_a Q_b = Q_a \}
\end{equation}
as the collection of 'local' probability measures that have compatible marginals with respect to the projections $i^a_b : E_a \to E_b$ when $b \subseteq a$ in $\mathcal{P}(I)$.
The next proposition expresses the free energy of probability distributions that factor according to an acyclic graph as a weighted sum of free energies of its marginals on edges and vertices. The weights are given by the inclusion-exclusion formula of the graph when seen as a poset. First, let us recall the inclusion-exclusion formula for a poset.

\begin{defn}[Zeta operator of a poset]
Let $\A$ be a finite poset. We call the ‘zeta-operator' of a poset $\A$, denoted $\zeta$, the operator from $ \bigoplus_{a\in \A}\R\to \bigoplus_{a\in \A}\R$ defined as, for any $\lambda\in \bigoplus_{a\in \A}\R$ and any $a\in \A$,
\begin{equation}
\zeta(\lambda)(a)=\underset{b\leq a}{\sum} \lambda_b
\end{equation}

\end{defn}

\begin{prop}[Reformulation of Proposition 2 \cite{Rota}, Rota' 64]\label{mobius-inversion}
Let $\A$ be a finite poset. The zeta-operator of $\A$ is invertible. We will call its inverse the M\"obius inversion of $\A$, denoted $\mu$. Furthermore, there is a collection $(\mu(a,b); b,a\in \A \text{ s.t. } b\leq a)$ such that, for any $\lambda\in \bigoplus_{a\in \A}\R$ and $a\in \A$,

\begin{equation}
\mu(\lambda)(a)=\underset{b\leq a}{\sum} \mu(a,b) \lambda_b
\end{equation}

\end{prop}

We call the coefficient $(\mu(a,b), b,a \text{ s.t. } b\leq a)$ the  M\"obius coefficients of $\A$. In particular Proposition \ref{mobius-inversion} implies that, for any $b,a\in \A$ such that $b\leq a$,

\begin{align}
\sum_{c:b\leq c\leq a} \mu(a,c) &= 1[b=a]\\
\sum_{c:b\leq c\leq a} \mu(c,b) &= 1[b=a]
\end{align}

For a collection of values $\lambda_a\in \mathbb{R}$, $a\in \mathcal{A}$, we call the following expression $\sum_{a\in \mathcal{A}}\sum_{b\leq a} \mu(a,b) \lambda_b$ the inclusion-exclusion formula over a poset $\mathcal{A}$; this formula corresponds to the value one would attribute to a maximal element, denoted $1$, added to $\mathcal{A}$:

\begin{equation}
\lambda_{1}:=\sum_{a\in \A}\sum_{b\leq a} \mu(a,b) \lambda_b
\end{equation}

which can be rewritten as $\lambda_{1}=\sum_{a\in \A}[\sum_{b\geq a} \mu(b,a)] \lambda_a$. We will denote $c(a)=\sum_{b\leq a} \mu(a,b)$ the weighted coefficients

To find the classical inclusion-exclusion formula, consider $I$ a finite set; the poset $\mathcal{A}$ is $(\mathcal{P}(I),\supseteq)$ with the reversed order. The quantities $\lambda_a$, for $a \in \mathcal{A}$, are $\vert A_i \vert$ when $a=i \in I$, and represent the cardinality of the sets $A_i$, as well as the cardinality of all possible intersections $\vert \cap_{i\in a} A_i \vert$ when $a\subseteq I$. In this setting, the maximal element $1$ has a value $\lambda_1 = \vert \cup_{i\in I} A_i \vert$.

\begin{prop}\label{equivalence-Bethe-on-DAG}
Let $I$ be a finite set, and let $E = \prod_{i\in I} E_i$ be a product of finite sets. Consider a finite acyclic graph $G = (I,A)$. Let $(H_a: E_a \to \mathbb{R}, a \in \mathcal{A})$ be a collection of Hamiltonians (measurable maps) that respectively factor through the projection $i^a: E \to E_a$. The following map,

\begin{equation}
\begin{array}{ccccc}
\phi & : & [\ast, i]_{\mathcal{A}} & \to & \Fac_{\mathcal{A}(G)}\\
& & (Q_a, a \in \mathcal{A}(G)) & \mapsto & \frac{\prod_{e\in A} Q_e}{\prod_{v\in I} Q_v^{d(v)-1}}\\
\end{array}
\end{equation}

is a bijection; here $d(v)$ is the degree of node $v$. Furthermore,

\begin{equation}
\mathbb{E}_{\phi(Q)}\left[\sum_{e\in A} H_e + \sum_{v\in I} H_i\right] - S(\phi(Q)) = F\left(Q_a, a \in \mathcal{A}(G)\right)
\end{equation}

where

\begin{equation}
F\left(Q_a, a \in \mathcal{A}(G)\right) = \sum_{a \in \mathcal{A}(G)} c(a) \left[S(i^a_{*}Q) - \mathbb{E}_{i^a_{*}Q}[H_a]\right]
\end{equation}

\end{prop}

\begin{proof}

Let for any $p\in \p_{>0}(E)\cap \Fac_{\A(G)}$,

\begin{equation}
\begin{array}{ccccc}
\psi&:& \p(E)\cap \Fac_{\A(G)}& \to & \lim F\\
& & P&\mapsto &(\pi_{a,*} P,a\in \A(G))\\
\end{array}
\end{equation}

Then $\phi\psi=\id$ and $\psi\phi=\id$, furthermore for any $P\in \p(E)\cap \Fac_{\A(G)}$,

\begin{equation}
\mathbb{E}_{Q}[\sum_{a\in \A}H_a]=\mathbb{E}_{Q}[ \sum_{a\in \A}\sum_{b\leq a} \mu(a,b) H_a]= \sum_{a\in \A}\sum_{b\leq a} \mu(a,b) \mathbb{E}_{i^{a}_{*}Q}[H_a]
\end{equation}

and by Proposition \ref{factorisation-marginalisation-on-DAG}, 

\begin{equation}
S(Q)= \sum_{e\in A} S(i^{e}_{*}Q) -\sum_{v\in I} (d(v)-1)S(i^{v}_{*}Q)=\sum_{a\in \A(G)} c(a) S(i^{a}_{*}Q)
\end{equation}

which ends the proof.

\end{proof}

$F\left(Q_a, a \in \mathcal{A}(G)\right) = \sum_{a \in \mathcal{A}(G)} c(a) \left[S(i^a_{*}Q) - \mathbb{E}_{i^a_{*}Q}[H_a]\right]$ is called the Bethe free energy and depends only on the marginal distributions $i^a_{*}Q$. Proposition \ref{equivalence-Bethe-on-DAG} implies that variational inference on acyclic graphical models can be done over marginal distributions and still give the exact posterior. This remark is very important as, for a graph $(G,E)$, the state space of $(E_e, e \in A)$ is of size $\vert A\vert \times N^2$ with $N$ the maximal size of the $E_v, v \in I$, whereas the cardinality of $E$ is $O(N^{\vert I\vert})$. The collection of marginal distributions serves therefore as a compressed representation of the joint distribution. When $G$ is acyclic, the Belief Propagation algorithm finds the minima of the Bethe free energy exploiting the compressed reformulation of $\Fac_{\mathcal{A}(G)}$; the complexity of the algorithm is $O(\vert A\vert)$.

The Bethe free energy can be defined on any graph, even those that may contain cycles; in such cases, the equality between the Bethe free energy and the Gibbs free energy does not hold in general. The key idea is to replace the Gibbs free energy with the Bethe free energy in variational inference and solve the associated optimization problem, hoping that the true posterior will be correctly approximated.

We introduced the Bethe free energy in a similar fashion to \cite{PeltrePhD}. This also allows us to see how one can define the Bethe free energy for any factorization model. In this context, the Bethe free energy is called the Generalized Bethe free Energy or region-based approximation of free energy in \cite{yedidia_constructing_2005}. In the next section, we define the (Generalized) Belief Propagation algorithm for the Generalized Bethe free Energy.

\section{Background: Generalized Bethe free energy}\label{Generalized Bethe Free Energy}

The Generalized Bethe free energy is an approximation of the Gibbs free energy that generalizes the Bethe free energy \cite{yedidia_constructing_2005}. Consider the configuration space $E=\prod_{i\in I} E_i$ of finite sets over a finite set $I$. Let $\A\subseteq \mathcal{P}(I)$, and let $(H_a\in E_a\to \R, a\in \A)$ be a collection of (measurable) Hamiltonians. For $Q=(Q_a\in \mathbb{P}(E_a), a\in \A)$, the General Bethe free energy is defined as:

\begin{equation}
F_{\text{Bethe}}(Q)=\sum_{a\in \A} \sum_{b\geq a} \mu(b,a) \left(\mathbb{E}_{Q_a}[H_a] - S_a(Q_a)\right)
\end{equation}

Here we considered $Q=(Q_a\in \mathbb{P}(E_a), a\in \A)$ to be any collection of distributions over the local variables $X_a\in E_a$; however, let's keep in mind that the collection we are interested in are $Q\in [\ast, i]_{\mathcal{A}}$ that are compatible with respect to marginalization.

In the Bethe Free energy, the entropy functional is replaced by a reconstruction of the 'local' entropies that only makes use of the entropy of the marginals. This is the term $S_{GB}(p)=\sum_{a\in \A} c(a) S(p_a)$ in $F_{\text{Bethe}}$. It is economical in the sense that it is a non-redundant way of computing entropy, similar to counting the cardinality of the union of sets using the Inclusion-Exclusion principle (see \cite{KuoTing} for a detailed presentation of this idea).

The General Belief Propagation is an algorithm that enables us to find the critical points of the Generalized Bethe Free Energy. A classical result states that fixed points of this algorithm correspond to critical points of that free energy. Let us now recall the expression of this algorithm.\\

For $\A \subseteq \mathcal{P}(I)$, $E$ a finite product of finite sets, and $(H_a, a \in \A)$ a collection of Hamiltonians. For two elements of $\A$, $a$ and $b$ such that $b \subseteq a$, two types of messages are considered: top-down messages $m_{a\to b} \in \R^{E_b}$ and bottom-up messages $n_{b\to a} \in \R^{E_a}$. The update rule is as follows: consider messages at times $t$, $(n_{b\to a}^t, m_{a\to b}^t \vert b, a \in \A \text{ s. t. } b \leq a)$, they are related by the following relation,

\begin{equation}\label{GBP1}
\forall a,b\in \A, \text{s.t. }  b\leq a, \quad n_{b \to a}^t= \prod_{\substack{c: b\leq c\\ c\nleq a}}m_{c\to b}^t
\end{equation}

One can define beliefs as ,

\begin{equation}\label{GBP2}
\forall a \in \A, \quad b_a^t\propto e^{-H_a}\prod_{\substack{b\in \A:\\ b\leq a}} n_{b\to a}^t
\end{equation}

The beliefs are sometimes interpreted as probability distributions up to a multiplicative constant; here to make the presentation clearer, we require that $b_a$ is a probability distribution. The update rule is given by,

\begin{equation}\label{GBP3}
\forall a,b\in \A, \text{s.t.} b\leq a\quad m_{a\to b}^{t+1}(x_b)= m_{a\to b}^{t}(x_b)\frac{\sum_{y_a: \pi^a_b(y_a)=x_b} b_a^t(y_a)}{b_b^t(x_b)}
\end{equation}

The multiplication of function $n_{b\to a}$ that have different domains is made possible because there is an the embedding of $\R^{E_b}$ into $\R^{E_a}$ implicitly implied in the last equation. 

The algorithm can be rewritten in a more condensed manner, updating only the top-down messages, for all $a,b\in \A$, such that $b\leq a$,

\begin{align}
 m_{a\to b}^{t+1}(x_b)&=  m_{a\to b}^t(x_b)\frac{\sum_{\substack{y_a: i^a_b(y_a)=x_b}} e^{-H_a(y_a)} \prod_{\substack{c\in \A:\\ c\subseteq a}}\prod_{\substack{d: c\subseteq  d\\ d\not \subseteq a}}m_{d\to c}^t (x_c)}{e^{-H_b(x_b)}\prod_{\substack{c\in \A:\\ c\subseteq b}}\prod_{\substack{d: c\subseteq  d\\ d\not \subseteq b}}m_{d\to c}^t (x_c)}
\end{align}

We will denote this update rule as $GBP$, $m(t+1)= GBP(m(t))$

\begin{thm}[Yedidia, Freeman, Weiss, Peltre]
Let $(m_{a\to b},a,b\in \A \text{s.t. } b\subseteq a)$ be a fix point of the Generalized Belief Propagation, i.e.

\begin{equation}
m= GBP(m)
\end{equation}

Let $(b_a,a\in \A)$ be the associated beliefs and let, for $a\in \A$, $p_a= b_a/ \sum_{x\in E_a} b_a(x)$ be the associated normalized beliefs. Then $(p_a,a\in \A)$ is a critical point of $F_{Bethe}$ under the constraint that $p\in \lim F$.

\end{thm}

\begin{proof}
Theorem 5.15\cite{PeltrePhD}, Theorem 5\cite{yedidia_constructing_2005}.

\end{proof}

\section{Entropy of $\A$-specifications and variational principle}

Let $\A$ be a finite poset and $\gamma=(G,F)$ be a specifications with $G(a)$ being a finite set for any $a\in \A$. We propose that the entropy of $Q\in \mathcal{G}_g(\gamma)$ to be $S_{GB}(Q)=\sum_{a\in \A} c(a) S(Q_a)$ and that the free energy is $F_{\text{Bethe}}(Q)=\sum_{a\in \A} c(a)\left(\mathbb{E}_{Q_a}[H_a] - S_a(Q_a)\right)$.

It might seem at first glance that it is sufficient to apply GBP to find the critical points of the free energy for specifications, but it is actually much trickier than that. In GBP the free energy is constrained over $Q\in [\ast, i]_{\mathcal{A}}$ but for $Q\in \mathcal{G}_g(\gamma)$, $Q\in [\ast, F]_{K,\A} $. The first significant difference is that for GBP the presheaf is prescribed to the one associated with marginalization whereas here $F$ can be anything. In Section 3.2 \cite{sergeantperthuis2022regionalized} (version 1) the case where $F$ is any presheaf is treated in detail and corresponds to a fair amount of generalization with respect to GBP as presented in \cite{yedidia_constructing_2005,PeltrePhD}. The second difference is that $F$ is a functor and not a presheaf; it is an essential difference tackled in the most general setting by Theorem 2.3 \cite{sergeantperthuis2022regionalized} (version 1). The intuition behind such a difference is as follows. The dual of the $F$ acts on Lagrange multipliers; the Lagrange multipliers are analogous to the messages `$m_{a\to b}$' of GBP. When $F$ is a presheaf its dual is a functor and one can send the Lagrange multipliers in a cell $F(a)$ but applying ${F^a_b}^{*}$ to $m_{a\to b}$ for any $b\leq a$ and so one can do the product of such messages (or the sum if one takes the logarithm). One cannot send the Lagrange multipliers for the $b$'s smaller than $a$ into the cell $F(a)^{*}$ when the $F$ is a functor. However, when there is $G$ such that $F$ is a section of $G$, $G$ can send the multipliers into $F(a)$ and one can build a message-passing algorithm that finds the critical points of the Bethe free energy for specifications. We now explicitly state Theorem 2.3 \cite{sergeantperthuis2022regionalized} (version 1) in the particular case of specifications and detail the message-passing algorithms for finding optimal Gibbs measures.\\

\textbf{Problem to solve:} The optimization problem we want to solve is the following, 

\begin{equation}
\inf_{Q\in \mathcal{G}_g(\gamma)} F_{\text{Bethe}}(Q)
\end{equation}

We will need to generalize the $\zeta$ and Möbius inversion of a poset to the ones for functors and presheaves.

\begin{defn}[Mobius inversion associated to a functor]
Let $G: \A\to \Mod$ be a functor from a finite poset to the category of $R$-modules with $R$ a ring; let $\zeta_{G}: \bigoplus_{a\in \A} G(a)\to \bigoplus_{a\in \A} G(a)$ be such that for any $a\in \A$ and $v\in \bigoplus_{a\in \A} G(a)$,

\begin{equation}
\mu_{G}(v)(a)=\sum_{b\leq a} \mu(a,b) G^b_a(v_b)
\end{equation}

\end{defn}

\begin{prop}
Let $G: \A\to \Mod$ be a functor from a finite poset to the category of modules, $\mu_{G}$ is invertible and its inverse, denoted $\zeta_{G}$, is defined as follows, for any $a\in \A$ and $v\in \bigoplus_{a\in \A} G(a)$,

\begin{equation}
\zeta_{G}(v)(a)=\sum_{b\leq a} G^b_a(v_b)
\end{equation}
\end{prop}

\begin{proof}
Let $v\in \bigoplus_{a\in \A} G(a)$ and $a\in \A$,

\begin{equation}
\zeta_{G}\mu_G(v)(a)= \sum_{b\leq a}\sum_{c\leq b} \mu(b,c) G^b_aG^c_b(v_c)
\end{equation}

therefore,

\begin{equation}
\zeta_{G} \mu_{G}(v)(a)= \sum_{c\leq a}\left(\sum_{b: \ c\leq b\leq a} \mu(b,c)\right) G^c_a(v_c)= G^a_a(v_a)
\end{equation}

Furthermore,

\begin{equation}
\mu_{G}\zeta_{G}(v)(a)= \sum_{b\leq a} \mu(a,b)\sum_{c\leq b} G^c_a(v_c)= v_a
\end{equation}

\end{proof}

\begin{rem}\label{chapitre-8:change-order-poset}
Let us remark for any poset $(\A,\leq)$ one can reverse the relations, in other words, for any $a,b\in \A$, 

\begin{equation}
a \leq_{op} b\iff b\leq a
\end{equation}

We sall also denote $\leq_{op}$ as $\geq$ and the corresponding poset as $\A^{op}$ or $(\A,\geq)$. One has that, for any $a,b\in \A$ such that $a\geq b$,

\begin{align}
\zeta_{\A^{op}}(b,a)= \zeta_{\A}(a,b)\\
\mu_{\A^{op}}(b,a)= \mu_{\A}(a,b)
\end{align}

In particular for any $G: \A\to \Vect$ functor from a finite poset to the category of modules,

\begin{equation}
\mu_{G^{*}}= (\mu_G)^{*}
\end{equation}

as for any $(l_a\in G(a)^{*},a\in \A)$,
\begin{equation}
\sum_{a\in \A} \sum_{b\leq a} \mu(a,b) l_aG^b_a= \sum_{b\in \A} \sum_{a\geq b} \mu(a,b) {G^{*}}^a_b(l_a)
\end{equation}
\end{rem}

In what follows, we go back to the convention for specification, which is that $G$ is a presheaf (to the category of measurable spaces). Let us define the function $FE(Q):\prod_{a\in\A}\mathbb{P}(E_a)\to \prod_{a\in \A} \R$ as $FE(Q)= (\mathbb{E}_{Q_a}[H_a]-S_a(Q_a),\,a\in \A)$, which sends a collection of probability measures over $\A$ to their Gibbs free energies. For any $Q\in \prod_{a\in\A}\mathbb{P}(E_a)$, let us denote $\mathrm{d}_Q FE$ as the differential of $FE$ at the point $Q$.

\begin{thm}

Let $\A$ be a finite poset, let $\gamma=(G,F)$ be a $\A$-specification such that $G(a)$ is a finite set for any $a\in \A$. Let $H_a:G(a)\to \R$ be a collection of (measurable) Hamiltonians. The critical points of the Generalized Bethe free energy are the $Q\in [\ast, F]_{K,\A} $ such that,

\begin{equation}
\mu_{G^*} \dd_Q FE|_{ [\ast, F]_{K,\A}}=0 
\end{equation}
\end{thm}

Let us now present the message-passing algorithm we consider. For two elements of $\A$, $a$ and $b$ such that $b\leq a$, two types of messages are considered: top-down messages $m_{a\to b}\in \R^{G(a)}$ and bottom-up messages $n_{b\to a}\in \R^{G(a)}$. Consider messages at times $t$, $(n_{b\to a}^t, m_{a\to b}^t ; b,a\in \A \text{ s.t. } b\leq a)$, they are related by the following relation,

\begin{align}
\forall b\leq a, \forall \omega_1\in G(a)\quad n^t_{b\to a}(\omega_1)=& \prod_{\substack{c: b\leq c\\ c\nleq a}} \sum_{\omega\in G(b)} m^t_{c\to b}(\omega) G^a_b(\omega\vert \omega_1)
\end{align}

One then defines beliefs as ,

\begin{equation}
\forall a \in \A, \quad b_a^t\propto e^{-H_a}\prod_{\substack{b\in \A:\\ b\leq a}} n_{b\to a}^t
\end{equation}

where $b_a\in \p(G(a)$. The update rule is,

\begin{align}
\forall a,b\in \A, \text{s.t.} b\leq a\quad m_{a\to b}^{t+1}&= m_{a\to b}^{t}\frac{F^b_a b_b^t}{b_a^t}
\end{align}

Theorem 2.2 \cite{sergeantperthuis2022regionalized} (version 2) applied to specifications implies that fixed points of the previous message-passing algorithm are in correspondence with critical points of the Bethe free energy over the space of Gibbs measures.

\section{Acknowledgement}

We would like to thank the reviewers for their comments. 
\bibliographystyle{ieeetr}
\bibliography{SYCO-bib}

\begin{thebibliography}{10}

\bibitem{sergeant-perthuis2024compositional}
G.~Sergeant-Perthuis, ``Compositional statistical mechanics, entropy and variational inference,'' in {\em Twelfth Symposium on Compositional Structures}, 2024.

\bibitem{sergeantperthuis2021bayesiangraphoid}
G.~Sergeant-Perthuis, ``Bayesian/graphoid intersection property for factorisation spaces,'' 2021.
\newblock \url{https://arxiv.org/abs/1903.06026}.

\bibitem{sergeantperthuis2021intersectiond}
G.~Sergeant-Perthuis, ``{Intersection property and interaction decomposition},'' Apr. 2019.
\newblock \url{https://arxiv.org/abs/1904.09017}.

\bibitem{sergeantperthuis2021interactionp}
G.~Sergeant-Perthuis, ``Interaction decomposition for presheaves,'' Aug. 2020.
\newblock \url{https://arxiv.org/abs/2008.09029}.

\bibitem{bennequin2020extrafine}
D.~Bennequin, O.~Peltre, G.~Sergeant-Perthuis, and J.~P. Vigneaux, ``Extra-fine sheaves and interaction decompositions,'' Sept. 2020.
\newblock \url{https://arxiv.org/abs/2009.12646}.

\bibitem{sergeantperthuis2021interaction}
G.~Sergeant-Perthuis, ``{Interaction decomposition for Hilbert spaces},'' 2021.
\newblock \url{https://arxiv.org/abs/2107.06444}.

\bibitem{GSPthese}
G.~Sergeant-Perthuis, {\em Intersection property, interaction decomposition, regionalized optimization and applications.}
\newblock PhD thesis, Université de Paris, 2021.
\newblock \href{https://www.researchgate.net/publication/349732943_Intersection_property_interaction_decomposition_regionalized_optimization_and_applications}{Link}.

\bibitem{10.1007/978-3-031-38271-0_26}
G.~Sergeant-Perthuis, ``A categorical approach to statistical mechanics,'' in {\em Geometric Science of Information} (F.~Nielsen and F.~Barbaresco, eds.), (Cham), pp.~258--267, Springer Nature Switzerland, 2023.

\bibitem{sergeantperthuis:hal-04456412}
G.~Sergeant-Perthuis, ``{Characterization of extreme Gibbs measures for a Categorical Approach to Statistical Mechanics},'' Feb. 2024.
\newblock \url{https://hal.sorbonne-universite.fr/hal-04456412}.

\bibitem{sergeantperthuis2022regionalized}
G.~Sergeant-Perthuis, ``Regionalized optimization,'' 2022.
\newblock \url{https://arxiv.org/abs/2201.11876}.

\bibitem{PeltrePhD}
O.~Peltre, ``Message passing algorithms and homology,'' 2020.
\newblock Ph.D. thesis, \href{https://opeltre.github.io/assets/bib/Peltre-Message_Passing_Algorithms_and_Homology.pdf}{Link}.

\bibitem{yedidia_constructing_2005}
J.~Yedidia, W.~Freeman, and Y.~Weiss, ``Constructing {Free}-{Energy} {Approximations} and {Generalized} {Belief} {Propagation} {Algorithms},'' {\em IEEE Transactions on Information Theory}, vol.~51, pp.~2282--2312, July 2005.

\bibitem{doi:10.1142/4090}
D.~Ruelle, {\em Statistical Mechanics}.
\newblock Imperial College Press, 1999.

\bibitem{928a56b7d6f1473e930f282a0c4b534e}
Y.~Lecun, S.~Chopra, R.~Hadsell, M.~Ranzato, and F.~Huang, {\em A tutorial on energy-based learning}.
\newblock MIT Press, 2006.

\bibitem{doi:10.1142/p579}
T.~Lelièvre, M.~Rousset, and G.~Stoltz, {\em Free Energy Computations}.
\newblock Imperial College Press, 2010.

\bibitem{chipot_free_2007}
C.~Chipot and A.~Pohorille, eds., {\em Free energy calculations: theory and applications in chemistry and biology}.
\newblock No.~86 in Springer series in chemical physics, Berlin ; New York: Springer, 2007.
\newblock OCLC: ocm79447449.

\bibitem{wiering_reinforcement_2012}
M.~Wiering and M.~van Otterlo, eds., {\em Reinforcement {Learning}: {State}-of-the-{Art}}, vol.~12 of {\em Adaptation, {Learning}, and {Optimization}}.
\newblock Berlin, Heidelberg: Springer Berlin Heidelberg, 2012.

\bibitem{Mezard}
M.~Mezard and A.~Montanari, {\em Information, Physics, and Computation}.
\newblock USA: Oxford University Press, Inc., 2009.

\bibitem{Gromov}
M.~Gromov, ``In a search for a structure, part 1: On entropy.,'' 2013.

\bibitem{Fritz}
T.~Fritz, ``A synthetic approach to markov kernels, conditional independence and theorems on sufficient statistics,'' {\em Advances in Mathematics}, 2020.

\bibitem{Vigneaux2020information}
J.~P. Vigneaux, ``Information structures and their cohomology,'' {\em Theory and Applications of Categories}, vol.~35, no.~38, pp.~1476--1529, 2020.

\bibitem{Spivak}
T.~Fritz and D.~I. Spivak, ``Internal probability valuations.'' Workshop Categorical Probability and Statistics, 2020.

\bibitem{Belfiore2021ToposAS}
J.-C. Belfiore and D.~Bennequin, ``Topos and stacks of deep neural networks,'' {\em ArXiv}, 2021.

\bibitem{Fritz1}
T.~Fritz and P.~Perrone, ``A probability monad as the colimit of spaces of finite samples,'' {\em Theory and Applications of Categories}, 2019.

\bibitem{Marcolli}
M.~Marcolli, ``Gamma spaces and information,'' {\em Journal of Geometry and Physics}, 2019.

\bibitem{Baudot2015}
P.~Baudot and D.~Bennequin, ``The homological nature of entropy,'' {\em Entropy}, vol.~17, no.~5, pp.~3253--3318, 2015.

\bibitem{fritz2021finetti}
T.~Fritz, T.~Gonda, and P.~Perrone, ``De finetti's theorem in categorical probability,'' {\em arXiv preprint arXiv:2105.02639}, 2021.

\bibitem{MOSS_PERRONE_2023}
S.~Moss and P.~Perrone, ``A category-theoretic proof of the ergodic decomposition theorem,'' {\em Ergodic Theory and Dynamical Systems}, vol.~43, no.~12, p.~4166–4192, 2023.

\bibitem{Georgii+2011}
H.-O. Georgii, {\em Gibbs Measures and Phase Transitions}.
\newblock Berlin, New York: De Gruyter, 2011.

\bibitem{alquier_approximate_2020}
P.~Alquier, ``Approximate {Bayesian} {Inference},'' {\em Entropy}, vol.~22, p.~1272, Nov. 2020.

\bibitem{PhysRev.65.117}
L.~Onsager, ``Crystal statistics. i. a two-dimensional model with an order-disorder transition,'' {\em Phys. Rev.}, vol.~65, pp.~117--149, Feb 1944.

\bibitem{PhysRevD.86.025022}
S.~El-Showk, M.~F. Paulos, D.~Poland, S.~Rychkov, D.~Simmons-Duffin, and A.~Vichi, ``Solving the 3d ising model with the conformal bootstrap,'' {\em Phys. Rev. D}, vol.~86, p.~025022, Jul 2012.

\bibitem{Simson1993LinearRO}
D.~Simson, ``Linear representations of partially ordered sets and vector space categories,'' 1993.

\bibitem{https://doi.org/10.48550/arxiv.2010.09651}
C.-S. Hu, ``A brief note for sheaf structures on posets,'' 2020.

\bibitem{YANAGAWA2001341}
K.~Yanagawa, ``Sheaves on finite posets and modules over normal semigroup rings,'' {\em Journal of Pure and Applied Algebra}, vol.~161, no.~3, pp.~341--366, 2001.

\bibitem{https://doi.org/10.48550/arxiv.2112.02609}
A.~Brown and O.~Draganov, ``Computing minimal injective resolutions of sheaves on finite posets,'' 2021.

\bibitem{incidence}
E.~Spiegel and C.~J. O'Donnell, {\em Incidence Algebras}.
\newblock CRC Press, 1997.

\bibitem{Giry}
M.~Giry, ``A categorical approach to probability theory,'' in {\em Categorical aspects of topology and analysis}, vol.~915 of {\em Lecture notes in Mathematics}, pp.~68--85, Springer, 1982.

\bibitem{Lawvere}
E.~W. Lawvere, ``The category of probabilistic mappings,'' 1962.
\newblock \href{https://ncatlab.org/nlab/files/lawvereprobability1962.pdf}{Link}.

\bibitem{mac2013categories}
S.~Mac~Lane, {\em Categories for the working mathematician}, vol.~5.
\newblock Springer Science \& Business Media, 2013.

\bibitem{gallier2022homology}
J.~Gallier and J.~Quaintance, {\em Homology, Cohomology, and Sheaf Cohomology for Algebraic Topology, Algebraic Geometry, and Differential Geometry}.
\newblock World Scientific, 2022.

\bibitem{Kesavan2009}
H.~K. Kesavan, {\em Jaynes' maximum entropy principle}, pp.~1779--1782.
\newblock Boston, MA: Springer US, 2009.

\bibitem{DEMARTINO2018e00596}
A.~{De Martino} and D.~{De Martino}, ``An introduction to the maximum entropy approach and its application to inference problems in biology,'' {\em Heliyon}, vol.~4, no.~4, p.~e00596, 2018.

\bibitem{10.5555/534975}
J.~Pearl, {\em Probabilistic Reasoning in Intelligent Systems: Networks of Plausible Inference}.
\newblock San Francisco, CA, USA: Morgan Kaufmann Publishers Inc., 1988.

\bibitem{lauritzen1996graphical}
S.~L. Lauritzen, {\em Graphical Models}.
\newblock Oxford Science Publications, 1996.

\bibitem{lecture2013probabilistic}
E.~Sudderth and J.~Pacheco, ``Probabilistic graphical models.'' Brown University CSCI 2950-P, Spring 2013, Lecture 12: Gaussian Belief Propagation, State Space Models and Kalman Filters, Guest Kalman Filter Lecture by Jason Pacheco, 2013.

\bibitem{doi:10.1143/JPSJ.12.753}
T.~Morita, ``Cluster variation method of cooperative phenomena and its generalization i,'' {\em Journal of the Physical Society of Japan}, vol.~12, no.~7, pp.~753--755, 1957.

\bibitem{10.1007/978-3-030-26980-7_23}
O.~Peltre, ``A homological approach to belief propagation and bethe approximations,'' in {\em Geometric Science of Information} (F.~Nielsen and F.~Barbaresco, eds.), (Cham), pp.~218--227, Springer International Publishing, 2019.

\bibitem{Speed}
T.~P. Speed, ``A note on nearest-neighbour gibbs and markov probabilities,'' {\em Sankhy\=a: The Indian Journal of Statistics, Series A}, 1979.

\bibitem{Rota}
G.-C. Rota, ``On the foundations of combinatorial theory {I}. {Theory} of {M}{\"o}bius functions,'' {\em Probability theory and related fields}, vol.~2, no.~4, pp.~340--368, 1964.

\bibitem{KuoTing}
H.~K.~T., ``On the amount of information,'' {\em Theory of Probability and its Applications}, 1962.

\end{thebibliography}

\end{document}